\documentclass[12pt,a4paper]{article}

\usepackage{latexsym}
\usepackage{graphicx}
\pdfoutput=1
\usepackage{pdfpages}   
\usepackage{color}
\usepackage{amsmath,amssymb}
\usepackage{cite,./mcite}
\usepackage{xspace}
\usepackage{ifthen}
\usepackage{hyperref}
\usepackage{slashed}
\usepackage[rightcaption]{sidecap}
\sidecaptionvpos{figure}{c}

\newcommand{\beq}{\begin{equation}}
\newcommand{\eeq}{\end{equation}}
\newcommand{\bea}{\begin{eqnarray}}
\newcommand{\eea}{\end{eqnarray}}

\def\vbfnlo{\textsc{Vbfnlo}}
\newcommand{\powheg}{\textsc{Powheg}}
\newcommand{\pbox}{\textsc{Powheg-Box}}
\newcommand{\herwig}{\textsc{Herwig++}}
\newcommand{\ds}{\textsc{DS++}}
\newcommand{\pythia}{\textsc{Pythia}}

\newcommand{\be}{\begin{eqnarray*}}
\newcommand{\ee}{\end{eqnarray*}}

\numberwithin{equation}{section}

\begin{document}

\thispagestyle{empty}

\begin{center}
\hfill KA-TP-04-2013
\end{center}

\begin{center}

\vspace{1.7cm}

{\LARGE 
Parton Shower Effects on $W$ and $Z$ Production \\[0.3cm] via Vector Boson Fusion at NLO QCD
}

\vspace{1.4cm}

{\bf Franziska Schissler
, \bf Dieter Zeppenfeld}\\ 

\vspace{1.2cm}

{\em {Institute for Theoretical Physics, Karlsruhe 
Institute of Technology, 76128 Karlsruhe, Germany}
}\\

\end{center}

\vfill

\centerline{\bf Abstract}
\vspace{2 mm}
\begin{quote}
\small
We present the implementation of electroweak $Zjj$ and $Wjj$ production via vector boson fusion with fully leptonic decays at NLO QCD in 
the \powheg \, framework. These processes represent an important background to Higgs searches in vector boson fusion, but they 
also can be seen as signal processes to study anomalous triple vector boson couplings as well as the impact of a central jet 
veto. Observables related to the third jet are sensitive to the parton shower which is used, a fact which is demonstrated by a 
comparison between \pythia, the 
standard angular-ordered \herwig \, shower and the new $p_T$-ordered Dipole Shower in \herwig.
\end{quote}

\vfill

\newpage
%
\section{Introduction \label{sec:Intro}}
With the discovery of a new boson at both \textsc{Atlas} \cite{AtlasHiggs} and \textsc{CMS} \cite{CMSHiggs} 
we have taken one step closer to the understanding 
of electroweak symmetry breaking. To achieve this goal, 
one needs to measure the couplings of this new boson to 
known Standard Model (SM) particles very precisely. Therefore, both signal and background processes must be understood in detail. 
In order to give reliable predictions for distributions at hadron colliders, it is mandatory 
to work at next-to-leading order (NLO) QCD and, if possible, interface this calculation with a parton shower 
to keep track of additional soft and collinear radiation. Up to now, there exist two prescriptions to match a NLO calculation 
to parton showers: \textsc{Mc@Nlo} \cite{MC@NLO}  and \powheg \cite{powheg}. 
We will focus on the latter. 
The \powheg \, method was implemented in a fully flexible program named \pbox \cite{powhegbox} which equips the users with 
all subroutines needed to match their fixed-order NLO calculation to a parton shower.\\

In case of the SM Higgs boson, 
 there already exists such a \pbox \, implementation for gluon fusion in association 
with zero jets \cite{ggH}, one and two jets \cite{Campbell:2012am} and vector boson fusion (VBF) \cite{Nason:2009ai}. 
VBF results in a very specific collider signature with one forward and one backward jet with a large 
rapidity gap between them which can be used to efficiently suppress background 
stemming from QCD-induced processes. To study this signature in data, one can look at a $W$ or $Z$ boson produced in 
VBF \cite{CMSZjj} and apply for instance central jet veto (CJV) techniques \cite{CJV} 
to this kind of processes. Since the cross section is higher for electroweak 
gauge boson production in VBF than for Higgs production, one can test the theoretical predictions for these processes before going 
to the real Higgs signal. 
Therefore, we implemented $Wjj$ and 
$Zjj$ production in VBF with subsequent leptonic decays in the \pbox \, to give a NLO prediction which can be interfaced 
with parton showers. The fixed-order $\alpha_s$ corrections to the cross section 
were already calculated in \cite{Oleari:2003tc} and we 
find good agreement with the already existing electroweak $Zjj$-implementation in 
the \pbox  \cite{Jager:2012xk}. The QCD induced $Zjj$ production is also 
part of the \pbox \cite{Re:2012zi} and can be used to test the efficiency of VBF cuts for background-suppression.\\

One goal of this work is to gain experience in interfacing an existing NLO code at fixed order in $\alpha_s$ with the \pbox. The processes 
explained in detail in this publication offer enough complexity to study the compatibility of \vbfnlo \cite{vbfnlo}, 
a fully flexible parton level Monte Carlo program for NLO QCD corrected cross sections and distributions, and 
the \pbox. The future plan is to make more processes implemented in \vbfnlo\, available in the \pbox. 
Additionally, we turn our attention to the influence of the parton shower on the studied processes. 
To this end, we study the $p_T$-ordered 
shower  in \pythia \cite{Pythia} as well as the vetoed, angular-ordered shower in \herwig \cite{HERWIG} and the new $p_T$-ordered
\herwig-Dipole Shower \cite{Platzer:2011bc}, in the following just called \ds. From these predictions we can  
estimate the influence of truncation to an angular ordered shower.\\
This paper is organized as follows: In Section \ref{sec:Impl} we review the details of the numerical calculation 
of all three processes, focusing on the subtleties of the matching between \vbfnlo \, and the \pbox. In Section \ref{sec:results} 
we will give results of our calculation, showered with \pythia, \herwig \, and \ds. 
Conclusions are given in Section \ref{sec:conclusions}.

\section{Details of the Implementation \label{sec:Impl}}
To interface the parton-level calculation of \vbfnlo \,  for $Zjj$ and $Wjj$ production via VBF with shower Monte-Carlo programs 
we use the publicly available \pbox \ framework \cite{powhegbox}. This package equips the user 
with all needed subroutines to go from a fixed-order NLO calculation 
in QCD to event files in the LesHouches format \cite{LesHouches} 
which then can be interfaced with a truncated shower. 
To this end, the \pbox \, asks for the following ingredients:
\begin{itemize}
 \item The Born squared matrix elements $\mathcal{B}$ for each partonic subprocess. The spin-correlated matrix elements 
        are not needed here since there are no external gluons at tree level. 
 \item The Born color structure in the limit of a large number of colors.
 \item The phase space for the Born process, see Section \ref{sec:ps}.
 \item The real emission squared matrix elements.
 \item The finite part of the interference term between the Born and virtual amplitude.
 \item The flavor structures of the Born and real emission subprocesses. We used tagging of the 
       different fermion lines as described in \cite{Nason:2009ai}. This means that same flavor fermions on 
       the upper and lower quark line internally get a different flavor (tag) to keep them distinct. These tags 
       are only used to assign the possible radiation regions, which are searched for automatically within the \pbox.
\end{itemize}
The details of the implementation of these ingredients will follow below. \\

The local subtraction terms needed to render the cross section finite are provided by the \pbox \, in the FKS framework \cite{FKS}. 
An automated check of all singular regions associated with one specific parton is performed and provides a good check for 
the flavor structures as well as the ratio between Born and real terms in the infrared (IR) region. Since the \pbox \, offers the 
possibility to generate fixed leading-order (LO) and NLO distributions with user-defined cuts, a cross check with fixed-order calculations 
performed with \vbfnlo \, is possible and provides a strong check for the validation of the implementation.\\

After these checks the \pbox \,  generates events in the LesHouches format from the \powheg-Sudakov factor which can be interfaced with any 
$p_T$-ordered shower like \pythia \, and \ds \, or to a truncated angular-ordered shower.
Since \herwig \, is an angular-ordered shower, one 
has to veto radiation harder than the real emission from the matrix element \cite{powheg}. 
This option is implemented in \herwig. However, one needs a so-called truncated shower 
to account for additional wide angle, soft radiation. 
This feature is not present in the current \herwig \, release. To estimate the effect of this additional soft radiation, we compare 
\herwig \, to the $p_T$-ordered \ds \, in our analysis.

\subsection{Matrix elements}
\begin{figure}[ht]
 \begin{center}
 \includegraphics[width=0.7\textwidth]{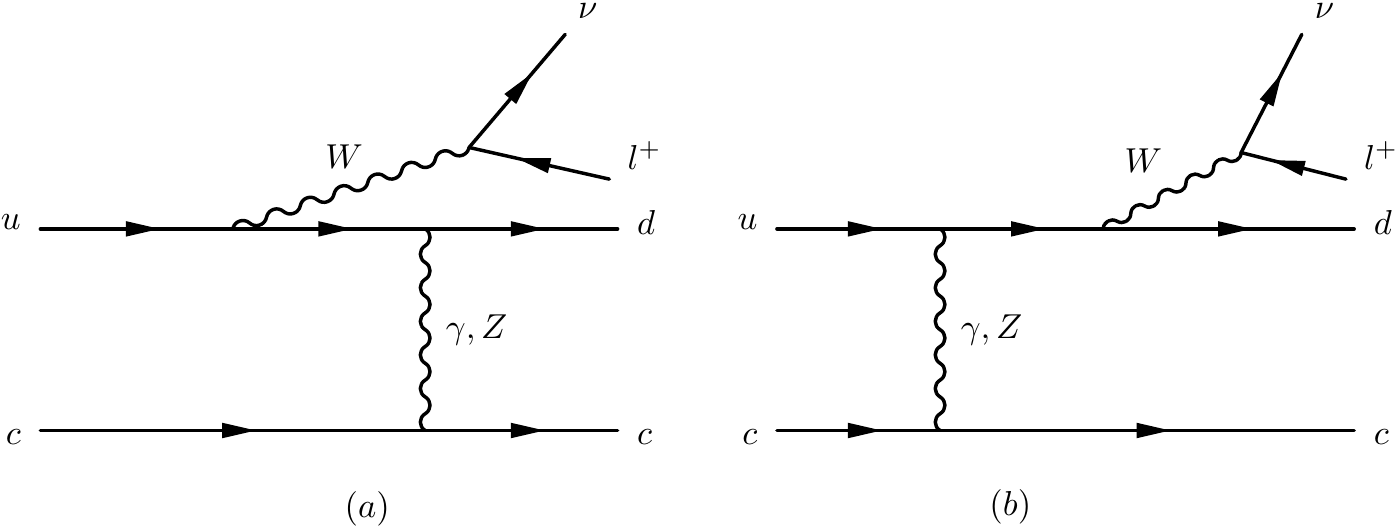} \\ \vspace{0.5cm}
 \includegraphics[width=0.7\textwidth]{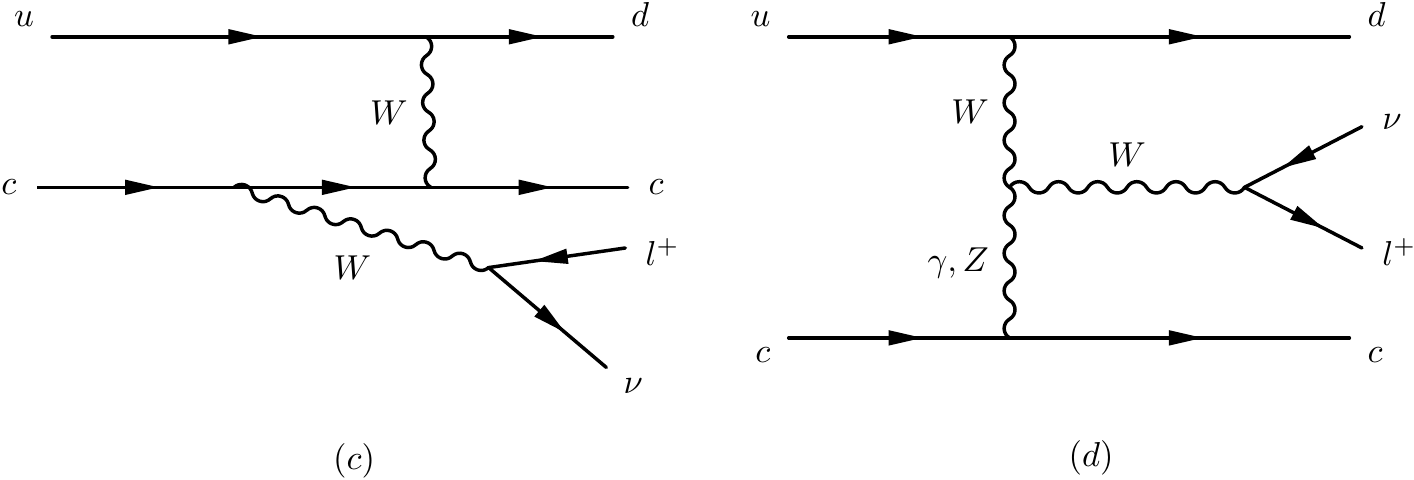} \\ \vspace{0.5cm}
 \includegraphics[width=0.7\textwidth]{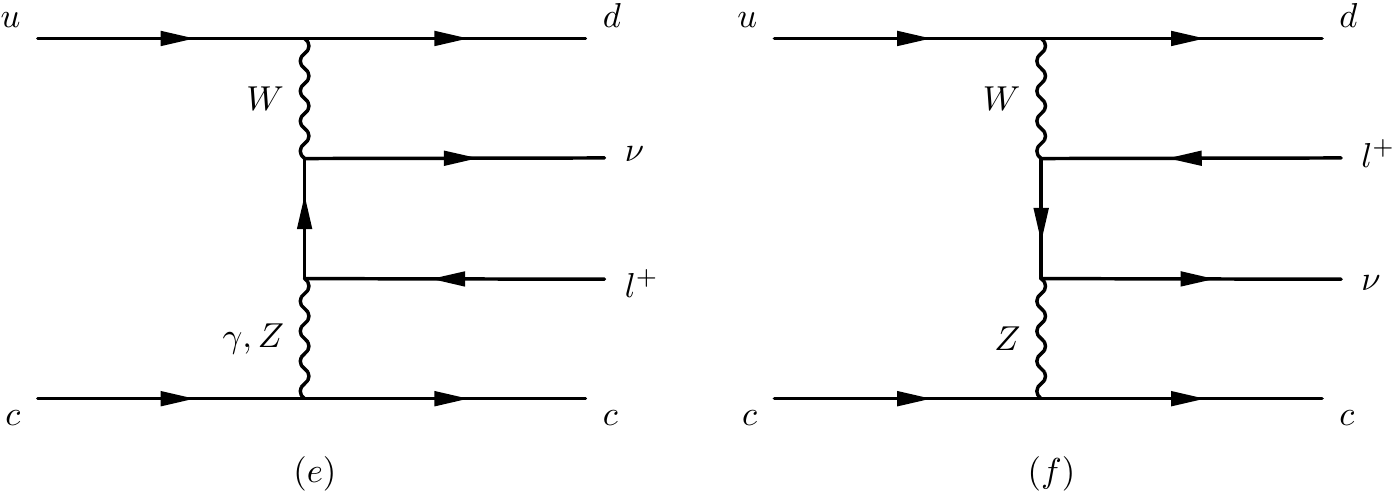}  
 \end{center}
 \caption{LO diagrams contributing to $Wjj$ production with subsequent leptonic decay. ($a$)-($d$) show resonant graphs, 
non-resonant graphs like ($e$) and ($f$) were also included.}
 \label{fig:born}
\end{figure}

\begin{figure}[t]
 \centering
 \includegraphics[width=0.7\textwidth]{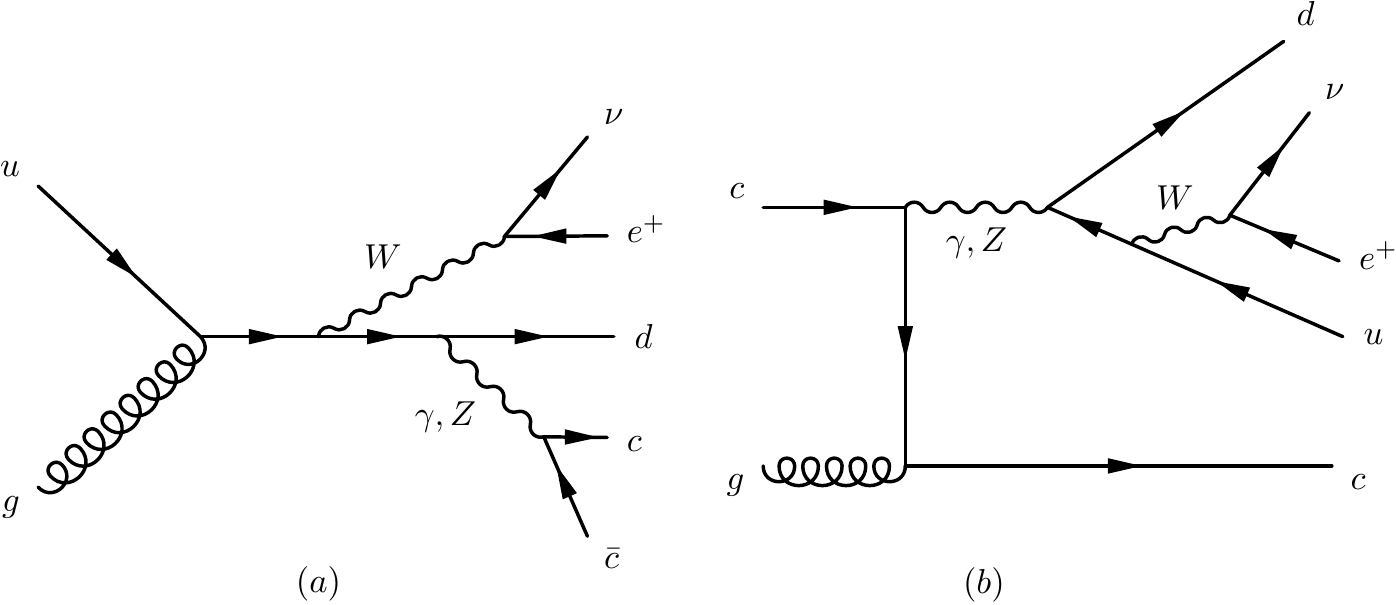}\\ \vspace{0.5cm}
 \includegraphics[width=0.7\textwidth]{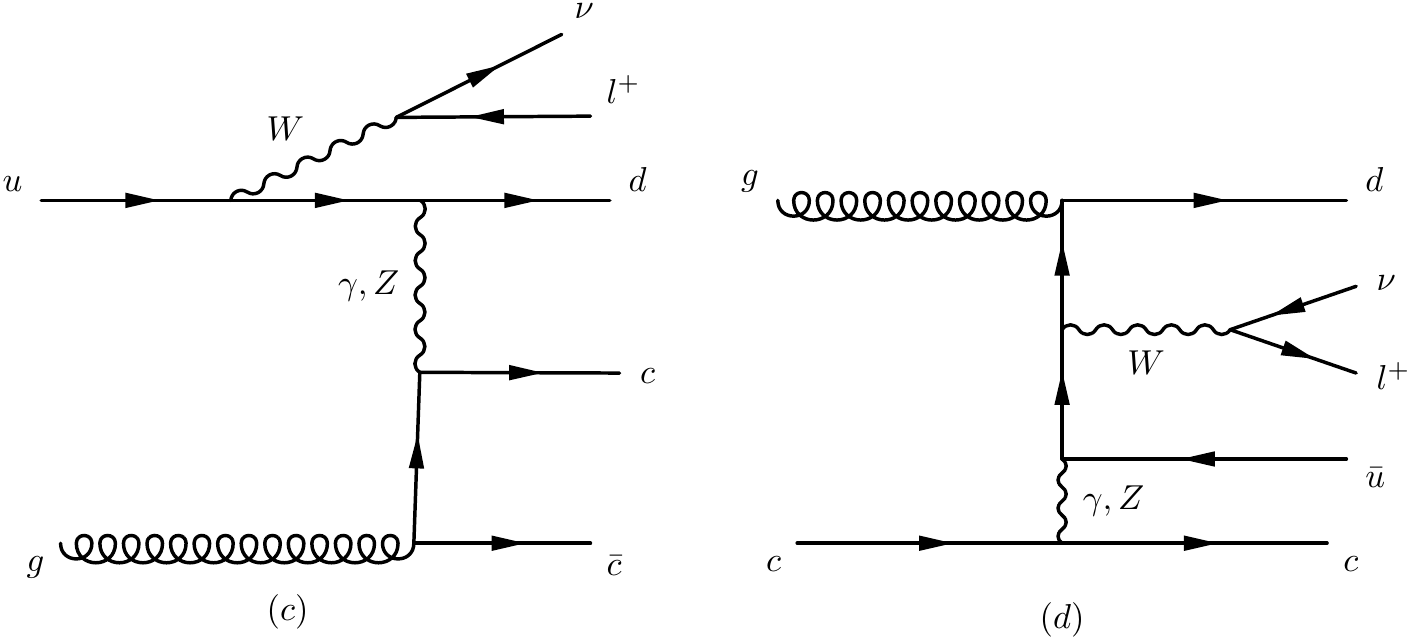}
 \caption{Real emission diagrams for $W$ production via VBF: 
($a$) and ($b$) are strongly suppressed in the VBF region and can be omitted, ($c$) and ($d$) 
show sample diagrams for gluon-induced amplitudes. Diagrams with 
final state gluons are obtained by crossing.}
 \label{fig:real}
\end{figure}

The matrix elements were adopted from the \vbfnlo \, implementation explained in detail in \cite{Oleari:2003tc}. 
Some sample diagrams for the Born and real emission 
contributions for $Wjj$-production are shown in Figures \ref{fig:born} and \ref{fig:real}. \\

When talking of $Vjj$-production ($V=W^\pm \text{ or } Z$), we mean on the one hand the 
resonant production of the vector boson with leptonic decay, where off-shell 
effects are fully taken into account through a modified version of the complex mass scheme \cite{Denner:1999gp} with real $\sin^2\theta_W$ 
and a Breit-Wigner integration of the propagator over the whole phase space. 
On the other hand, we also take non-resonant production of the leptons into account, see Figure \ref{fig:born} ($e$), ($f$). 
For $Zjj$-production, we also take a $\gamma^*$ with subsequent leptonic decay into account.
Fermion masses were set to zero throughout and $b$-quarks in 
the initial state were neglected. Also, the Cabibbo-Kobayashi-Maskawa matrix was set to the unit matrix. This is no approximation to the calculation as 
long as the flavor of the jets is not tagged and quark masses are neglected. \\

In Figure \ref{fig:real} ($a$) and ($b$), both (virtual) vector bosons are time-like and contain one vector boson which decays into a quark-anti-quark pair. 
This is a real emission contribution to $WZ$ production and treated as a separate process. This type of diagrams is therefore not considered here. 
More details on the 
used approximations can be found in \cite{Oleari:2003tc}.\\

One problem in the \pbox \, implementation is the presence of one ($Wjj$) or even two ($Zjj$) $t$-channel photons already at LO. Consequently, 
the Born cross section is divergent if integrated over the whole phase space, though, of course, 
it is well defined with normal jet definition cuts. To avoid these singularities, amplitudes which contain a $t$-channel 
photon with virtuality $Q^2< 4 \, \text{GeV}^2$ are suppressed with a large damping factor, as already used in \cite{Jager:2012xk}. 
At NLO, the $Q^2< 4 \, \text{GeV}^2$ cut affects the real emission contributions, as long as photon induced processes, 
which absorb the divergence via the photon pdf, are not taken into account. In the following
the missing $p\gamma \rightarrow VjjX$ piece is considered as a separate electroweak contribution to $Vjj$ production and, since these 
contributions are quite small when typical VBF cuts are imposed, they were neglected.

\subsection{Phase space \label{sec:ps}}
In our implementation, there exist three possibilities to evaluate the Born phase-space integral. 
The standard procedure maps the random numbers given by the integration routines to the physical momenta, adopted from \cite{Jager:2012xk}. 
Since the Born contributions are divergent in 
certain regions of the phase space, one can impose a cut on the two tagging jets, which can be changed 
by the user. The invariant dilepton mass for $Zjj$-production is required 
to be above 20 GeV by default, thus avoiding the $\gamma^*\rightarrow l^+l^-$ singularity 
at $Q^2=0$ for massless leptons.  \\

As already used in \cite{Jager:2012xk} and originally described in \cite{Alioli:2010qp}, 
there also exists the possibility to use a Born-suppression factor $F(\Phi_n)$ instead of the 
generation cut on the jets' $p_T$ described above. One possible choice for  $F(\Phi_n)$ is 
\beq
F(\Phi_n) = \left( \frac{p_{T,j_1}^2}{p_{T,j_1}^2+\Lambda_{p_{T_j}}^2} \right)^k \left( \frac{p_{T,j_2}^2}{p_{T,j_2}^2+\Lambda_{p_{T_j}}^2} \right)^k.
\eeq
This factor vanishes whenever a singular region in the Born phase space $\Phi_n$ is reached. The underlying Born kinematics are then generated 
in the \pbox \, according to a modified $\overline{B}$ function,
\beq
\overline{B}_{\text{supp}}=\overline{B}(\Phi_n) F(\Phi_n).
\eeq
The parameters $\Lambda_{p_{T_j}}=10$ GeV and $k=2$ can be changed by the user. The resulting events have to be reweighted by a factor $1/F(\Phi_n)$.\\

To speed up the generation of \powheg \, events it is also possible to use unweighted events generated by \vbfnlo\, 
as phase-space generator\footnote{This option can be used by setting the variable 
{\tt Phasespace} to 2 in the {\tt powheg.input} file.}. The main advantages of this approach is that, first of all, the integration over the Born 
variables and the optimization of the grid with respect to the underlying Born kinematics can be omitted, since the unweighted events are already 
flat in the (Born) phase space. These unweighted events can therefore 
be seen as the perfect LO phase-space generator. Only the integration over the 
three variables of the real emission has to be handled by the integration routine. To 
use this option, unweighted events were generated using \vbfnlo. Each 
event $i$ which survived the unweighting procedure was reweighted by the factor $$J_i=\frac{\sigma_{LO}}{\left(\left|\mathcal{M}_B\left(\Phi_n^{(i)}\right)\right|^2 \,\text{pdf}\left(\Phi_n^{(i)}\right)\right)},$$ 
the Born cross section over the respective numerical value of the squared Born matrix element including pdfs. 
This factor $J_i$ is exactly the Jacobi factor of the Born phase space. A Monte Carlo integration over $N$ reweighted events 
then reproduces the Born cross section:
\begin{eqnarray}
\frac{1}{N}\sum_{i=1}^N J_i \left|\mathcal{M}_B\left(\Phi_n^{(i)}\right)\right|^2\,\text{pdf}\left(\Phi_n^{(i)}\right) &=& 
\frac{1}{N} \sum_{i=1}^N \frac{\sigma_{LO}}{\left|\mathcal{M}_B\left(\Phi_n^{(i)}\right)\right|^2 \,\text{pdf}\left(\Phi_n^{(i)}\right)} \left|\mathcal{M}_B\left(\Phi_n^{(i)}\right)\right|^2 \,\text{pdf}\left(\Phi_n^{(i)}\right) \nonumber\\
&=& \sigma_{LO}.
\end{eqnarray}
For the numerical analysis shown in Section \ref{sec:results} we used this third method.

\subsection{Checks}
To check the implementation, all matrix elements were compared phase-space pointwise with the existing \vbfnlo \, subroutines. 
With this method, the evaluated couplings and modified routines were validated. 
Agreement from 11 to 15 digits was found. 
It was also verified that the subtraction and the real emission terms cancel in the singular limit.
Another important 
check is the agreement of differential fixed-order NLO distributions. All tested distributions agree between the \pbox\, and \vbfnlo\, implementation 
within statistical errors of at most 1 \%.
The validation of the use of unweighted events was done by comparing cross sections and distributions at fixed order and after event 
generation using the three different options to generate the phase space. Good agreement within the statistical errors was found.
For $Zjj$ production, we also compared our implementation to \cite{Jager:2012xk} using generation cuts in the phase space generator.
Matrix elements were compared phase-space pointwise and agreement at the level of 10 relevant digits was found. 
Cross sections and distributions agree within the 
statistical uncertainties at NLO and after event generation.

\section{Numerical Results \label{sec:results}}
For our numerical analysis for the LHC with center-of-mass energy of 8 TeV 
we operate with the CT10 pdf set \cite{Lai:2010vv} with $\alpha_s(M_Z)=0.11798$ as implemented in the \textsc{Lhapdf} package \cite{lhapdf}. 
For the calculation of the electroweak couplings we use the input parameters $M_W=80.398$ GeV, 
$M_Z=91.1876$ GeV and the Fermi constant $G_F=1.16637\cdot 10^{-5} \text{ GeV}^{-1}$. From these parameters the total widths of the 
electroweak gauge bosons are calculated to be $\Gamma_Z=2.5084\text{ GeV}$ and $\Gamma_W=2.0977\text{ GeV}$. The QED fine structure constant is 
$\alpha_{QED}=1/132.341$ and the weak mixing angle is $\sin^2 \theta_W=0.2226$.
Partons are recombined into jets according to the anti-$k_T$ algorithm \cite{Cacciari:2008gp} 
provided by the \textsc{FastJet}-package \cite{fastjet} with a default distance parameter $R=0.5$. \\

For the numerical analysis presented below we use the following inclusive cuts:\\
We require that the two highest $p_T$ jets, called tagging jets, satisfy 
\beq p_{T,j}^{tag}>30 \text{ GeV}. \label{eq:tag}\eeq
All observable jets, from the NLO calculation or the Shower, are demanded to have 
\beq p_{T,j}>20 \text{ GeV},\label{eq:3rd}\eeq
as well as rapidity 
\beq |y_j| < 4.5. \label{eq:yj}\eeq
To have well-observable leptons in the central region of the detector, they should obey
\beq  p_{T,l}>20 \text{ GeV} \qquad \text{and} \qquad |y_l| < 2.5.\label{eq:lep}\eeq
Since in $Zjj$-production, the process $\gamma^*jj\rightarrow  l^+ l^-jj$ is included as well, one is forced to impose a cut 
on the invariant mass of the leptons to avoid singularities:
\beq m_{ll}>20 \text{ GeV}.\label{eq:mll}\eeq
All leptons should be well separated from each other and from the jets, assured by
\beq \Delta R_{ll}>0.1 \qquad \text{and} \qquad \Delta R_{jl} > 0.4, \label{eq:jetlep}\eeq 
where $\Delta R_{ij}=\sqrt{(y_i-y_j)^2+(\phi_i-\phi_j)^2}$.\\

Due to the color singlet exchange in the $t$-channel, the two tagging jets are widely separated in rapidity and usually lie in opposite 
detector hemispheres. Additionally, the decay products of the weak boson tend to be located in the rapidity gap between the two tagging jets. 
This special configuration can be used to suppress QCD backgrounds, which have 
a higher jet activity in the central detector. We therefore demand the typical VBF-cuts
\begin{eqnarray} 
m_{jj}>600 \text{ GeV}, \quad \Delta y^{tag}_{jj} > 4,\quad y^{tag}_{j1} \times y^{tag}_{j2} <0 \quad
\text{and} \quad y^{min}_{j,tag}+0.2<y_l<y^{max}_{j,tag}-0.2.\nonumber \\
\label{eq:vbf}
\end{eqnarray}

The factorization and renormalization scale is set to the produced vector boson's mass $\mu_F=\mu_R=M_V$.\\

In the following we will discuss $W^+jj$ production with decay into the leptons of the first family. 
The main findings are the same for $W^-jj$ and  $Zjj$ production so only 
plots for the $W^+jj$ case will be shown. Since we are mostly interested 
in the effects of the three parton showers, \pythia, \herwig \, and \ds,
hadronisation and underlying event simulations were not taken into account. 
We used \pythia-version 6.4.25 with the Perugia 0-tune (Feb 2009) 
and \herwig-version 2.6.1a for the standard shower and for \ds. 
The cross sections with the VBF-cuts mentioned above are shown in \mbox{Table \ref{table:XSec}}.

\begin{table}[h]
\centering
  \begin{tabular}{|c||c|c|c|}
  \hline
    & $W^+jj$ & $W^-jj$ & $Zjj$ \\
   \hline
   NLO & $(253.9 \pm 0.3)$ fb & $(134.4 \pm 0.2)$ fb & $(24.47 \pm 0.07)$ fb \\
  \hline
   \textsc{Vbfnlo} & $(254.0 \pm 0.1)$ fb & $(134.6 \pm 0.1)$  fb & $(24.48 \pm 0.02)$ fb \\
  \hline
   \pythia & $(251.0 \pm 0.8)$ fb &  $(131.7 \pm 0.5)$ fb & $(24.48 \pm 0.18)$ fb\\
  \hline
   \herwig& $(249.8 \pm 0.8)$ fb &  $(131.2 \pm 0.5)$ fb& $(24.08 \pm 0.18)$ fb\\
  \hline
   \ds &  $( 245.2 \pm 0.8)$ fb & $(128.0 \pm 0.5)$ fb & $(23.56 \pm 0.18)$ fb\\
  \hline
  \end{tabular}
\caption{Cross sections for electroweak $Vjj$ production including VBF-cuts (\ref{eq:tag}-\ref{eq:vbf}) with subsequent decay of the vector boson into the first lepton family. The 
NLO cross section was obtained with the new \pbox \, implementation and matches the \vbfnlo \, prediction. \pythia \,
 and \textsc{Herwig} results include parton shower.}
\label{table:XSec}
\end{table}

\begin{figure}[t]
\begin{minipage}[l]{8cm}
	\centering
	\includegraphics[angle=-90,width=1\textwidth]{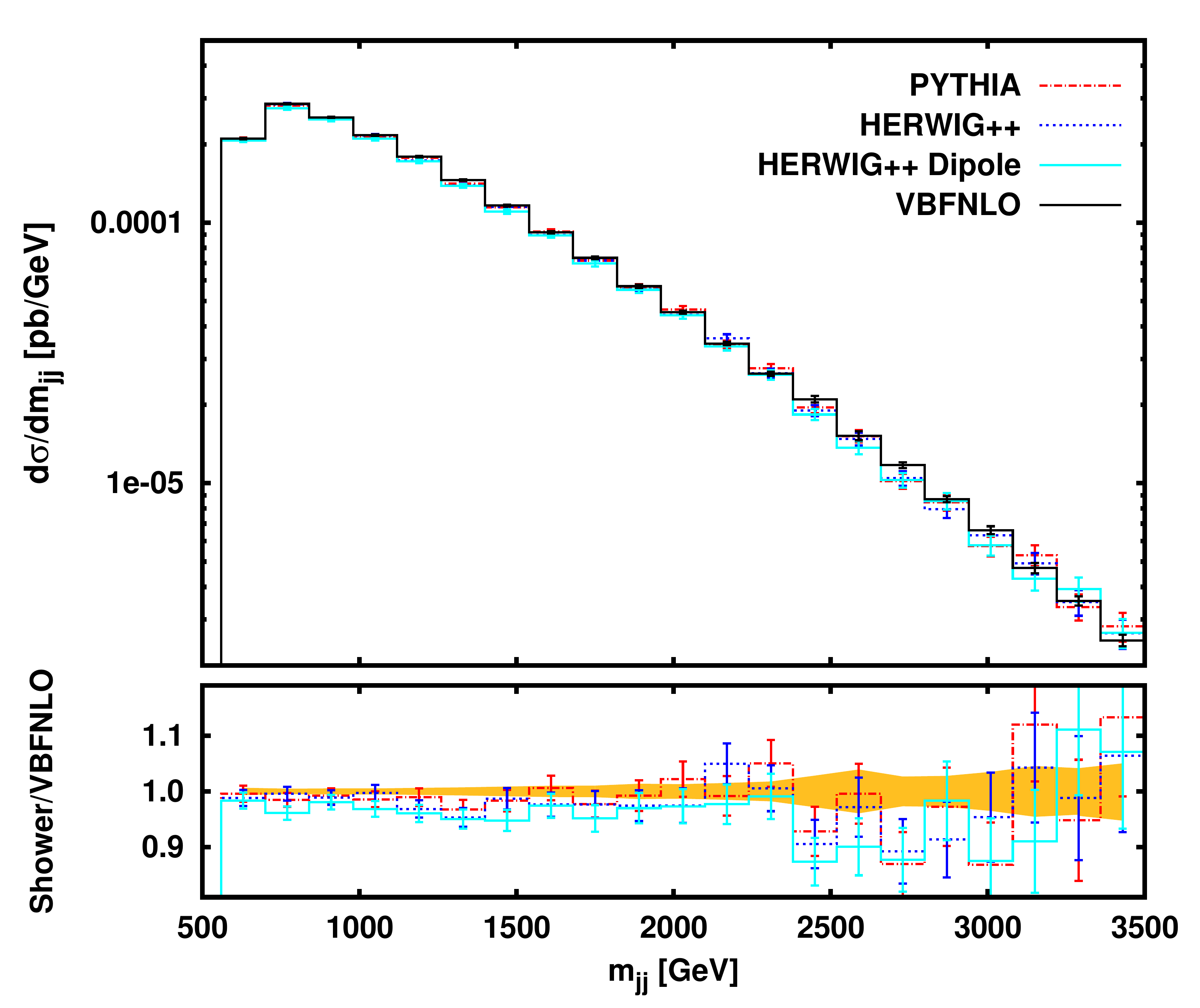}
\end{minipage} \begin{minipage}[r]{8cm}
	\centering
	\includegraphics[angle=-90,width=1\textwidth]{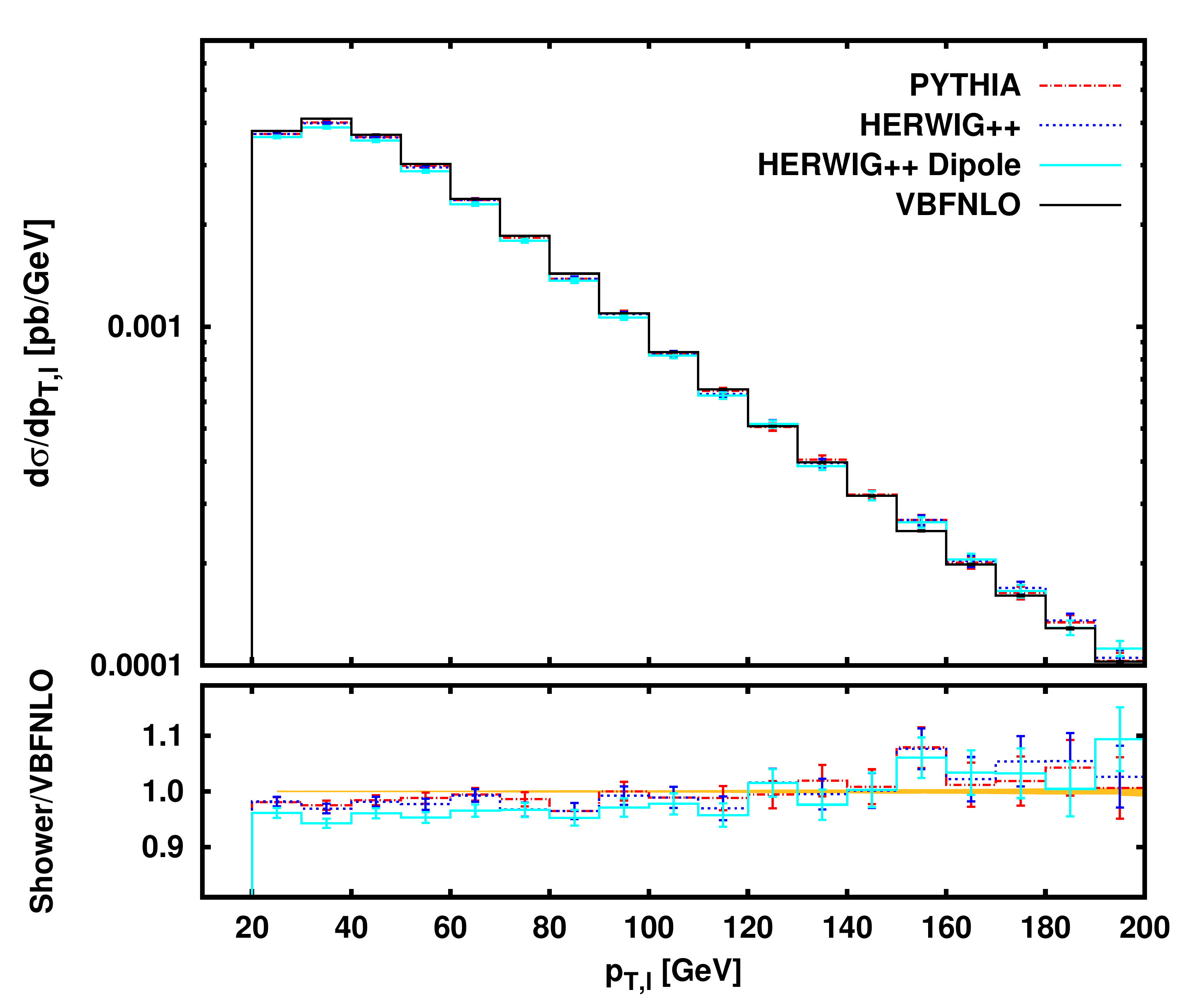}
\end{minipage}
\caption{Differential cross-section of the invariant tagging jet mass (left) 
and the transverse momentum of the charged lepton (right) of the \powheg-prediction compared to the fixed order 
curves of \vbfnlo \, (black solid line). The (red) dashed-dotted line shows the prediction of the \powheg \ result showered with 
\pythia, the (blue) dotted line corresponds to \herwig \, and the (turquoise) solid line to \ds. 
The error bars show the statistical error of the integration, the yellow 
error band in the ratio plot gives the statistical error on the fixed-order NLO result.}
\label{fig:same}
\end{figure}

Figure \ref{fig:same} shows the invariant mass of the two tagging jets and the transverse momentum of the charged lepton for
\pythia, \herwig \, and \ds \, in comparison to the fixed order NLO prediction of \vbfnlo. As expected, the parton showers have no effect 
on these observables except for a slight change in the normalization due to the different total cross sections. This change comes 
from events which pass the cuts in a fixed order calculation but migrate slightly by parton shower effects to phase space regions 
which are not incorporated within the cuts. Also, other observables constructed from the four-momenta of the tagging jets or the leptons are not affected by the parton showers, the VBF signature 
of the events is therefore preserved.\\

\begin{figure}[t]
\begin{minipage}[l]{8cm}
	\centering
	\includegraphics[angle=-90,width=1\textwidth]{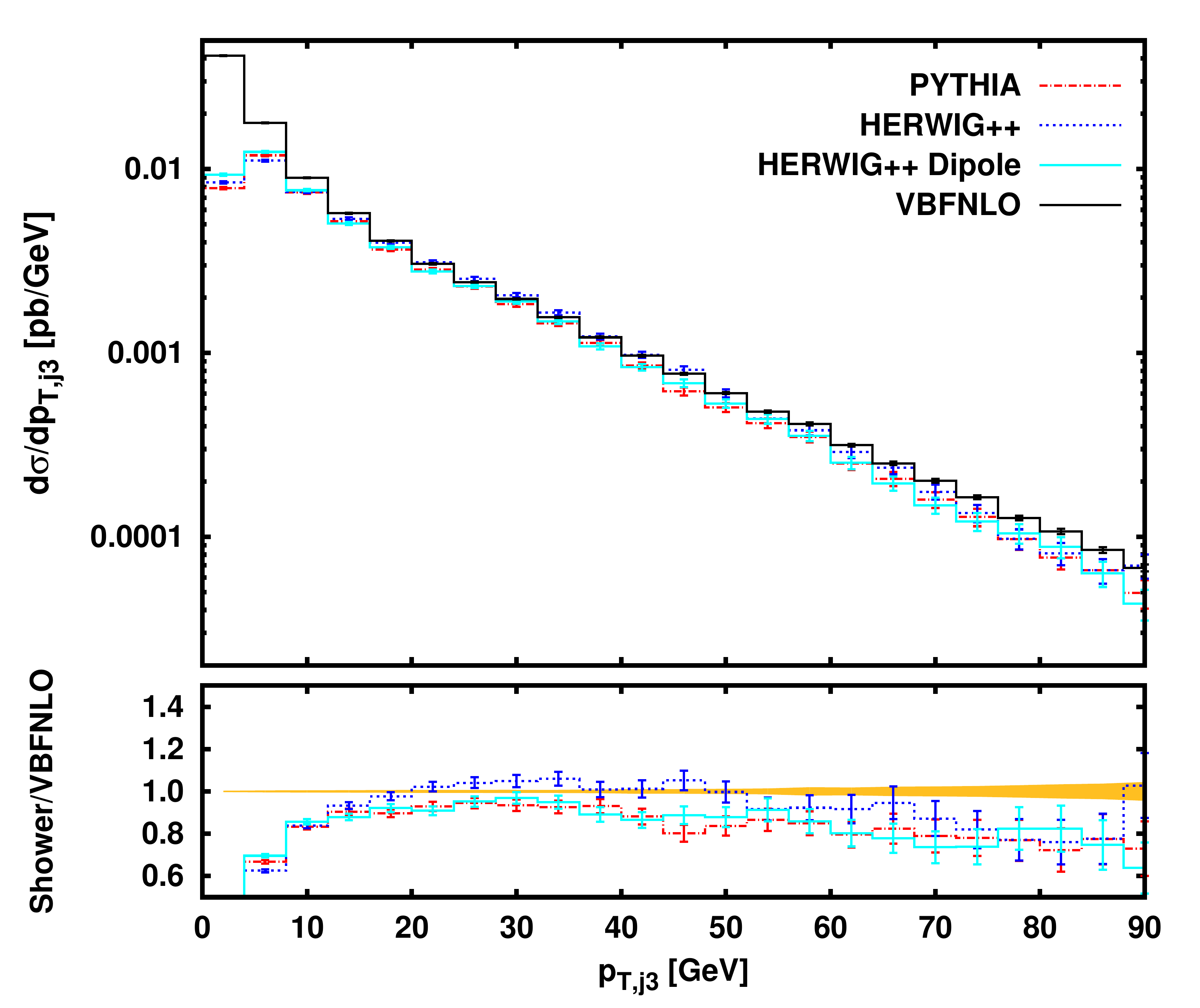}
\end{minipage} \begin{minipage}[r]{8cm}
	\centering
	\includegraphics[angle=-90,width=1\textwidth]{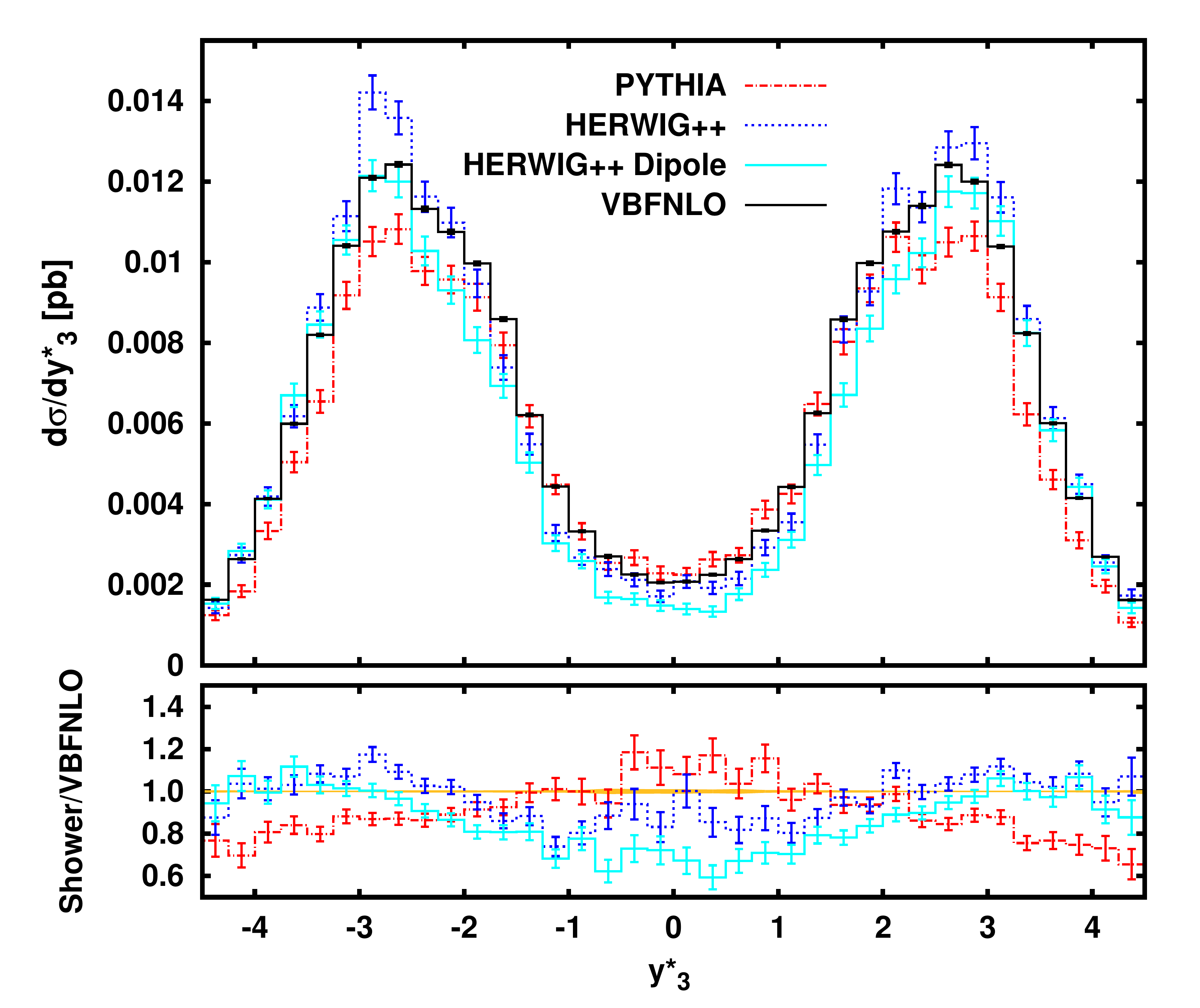}
\end{minipage}
\caption{Differential distribution of the transverse momentum of the third jet with $p_{T,j_3}>1 \text{ GeV}$ (left) 
and the variable $y_3^*$ (\protect\ref{eq:y3*}) (right) including all cuts, comparing the predictions of the three parton showers to the fixed order 
distributions of \vbfnlo. The line styles are assigned as in Figure \protect\ref{fig:same}.}
\label{fig:j3}
\end{figure}
Differences occur in the differential distributions of the third hardest jet, whose matrix elements are only LO accurate.
Figure \ref{fig:j3} shows the $p_T$-spectrum of the third jet and its location relative to the tagging jets, 
\beq 
y_3^*=y_{j_3} - (y_{j_1} + y_{j_2})/2. 
\label{eq:y3*}
\eeq
For the plot on the left hand side the 
cut on the transverse momentum of the third jet was lowered to $p_{T,j_3}>1$ GeV, 
whereas the right plot contains the usual VBF-cuts (\ref{eq:tag})-(\ref{eq:vbf}).\\

In the low $p_T$ region, the damping of the soft divergence due to the Sudakov factor can be observed 
for all three parton showers.
Between 20 and 50 GeV, \herwig \, predicts more jets than \pythia, but matches the NLO prediction, whereas \pythia \,
and \ds \, are in good agreement. In the tail of the 
distributions for hard jets with $p_T\gtrsim75 \text{ GeV}$, all three parton showers have lower rates than the NLO distribution. 
This comes from additional hard and/or wide angle radiation which can lead to additional jets which are not re-clustered in the direction 
of the parent parton.\\

Even bigger differences occur for the differential distribution of the variable $y_3^*$ (\ref{eq:y3*}), see the right plot of Figure \ref{fig:j3}.
With the $\Delta y^{tag}_{jj} > 4$ cut, the two tagging jets peak at $|y_j^{\text{tag}}| \approx 2.7$, so $|y_3^*|\lesssim 2.7$ typically corresponds to the rapidity gap between 
the tagging jets and $|y_3^*|\gtrsim 2.7$ to the third jet being positioned between the tagging jets and the beam axis. 
\pythia \, tends to radiate more into the rapidity gap and additionally underestimates the region between 
the tagging jets and the beam axis, whereas \herwig \, an \ds \, behave the opposite way. \\

\begin{figure}[t]
\begin{minipage}[l]{8cm}
	\centering
	\includegraphics[angle=-90,width=1\textwidth]{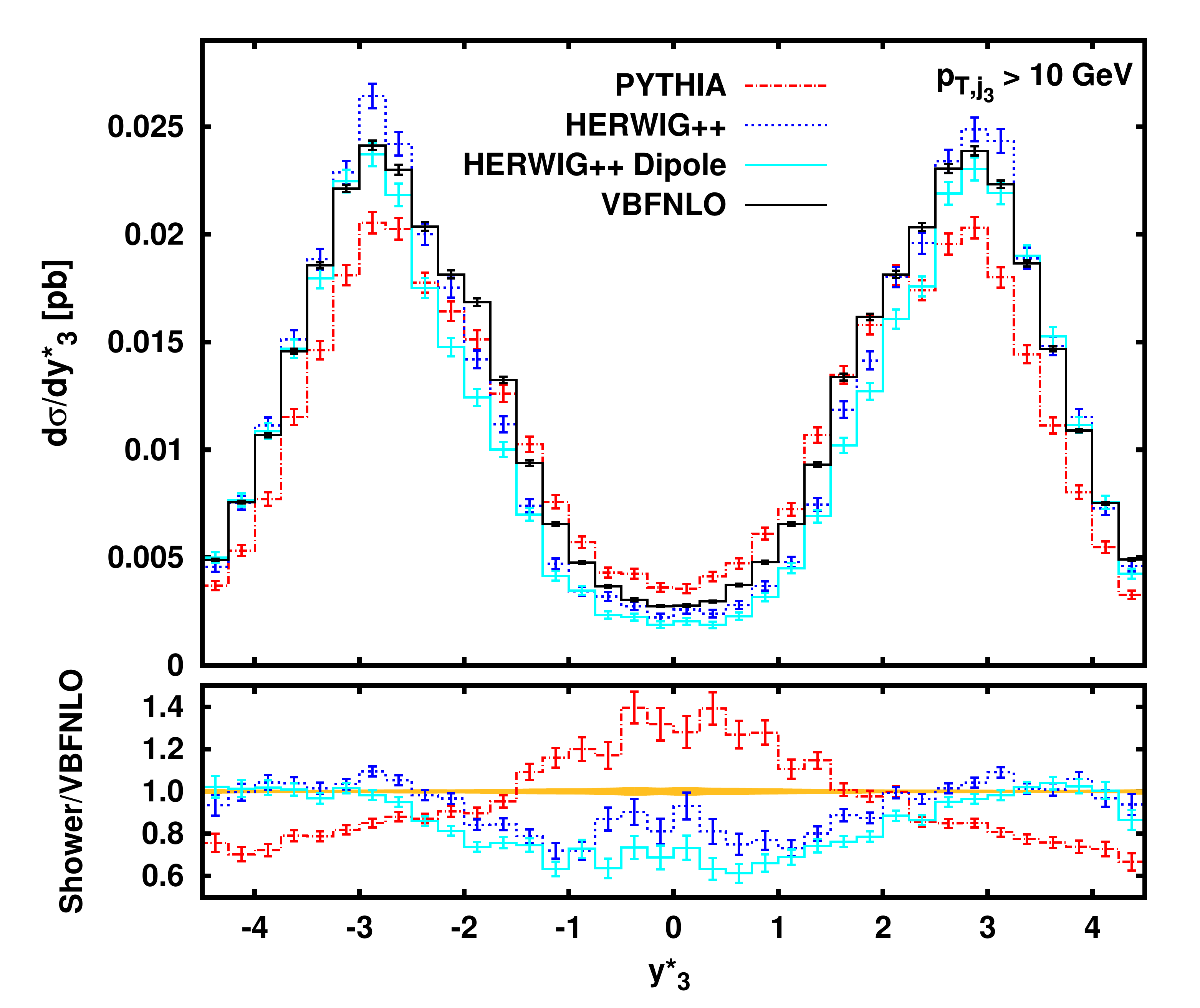}
\end{minipage} \begin{minipage}[r]{8cm}
	\centering
	\includegraphics[angle=-90,width=1\textwidth]{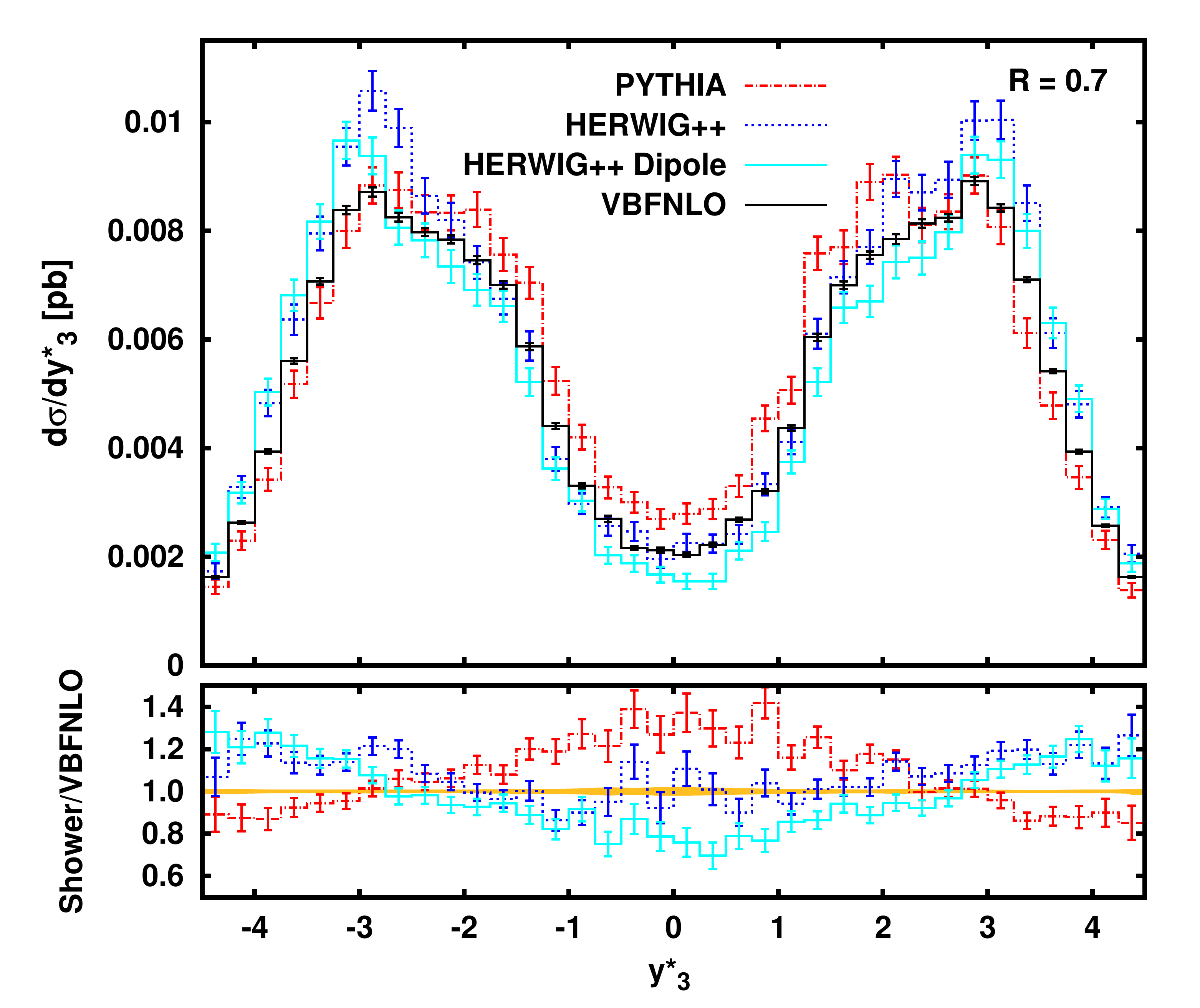}
\end{minipage}
\caption{Left: Differential $y_3^*$-distribution (\protect\ref{eq:y3*}) for $p_{T,j_3} > 10 \text{ GeV}$.
Right: Differential $y_3^*$-distribution for $R=0.7$. The line styles are assigned as in Figure \protect\ref{fig:same}. }
\label{fig:07}
\end{figure}

This effect gets even more pronounced 
if one varies the distance parameter $R$ of the anti-$k_T$ algorithm or lowers the $p_T$-cut on the third jet, see Figure \ref{fig:07}. 
The differences of the showers are due to the fact that \pythia \, tends to emit more soft partons, whereas \herwig \, and \ds \, preferentially emit 
partons in the collinear region between the tagging jets and the beam axis and therefore pull the third jet in that direction. 
If one lowers the $p_{T,j_3}$-cut, Figure \ref{fig:07} left, \pythia \, fills the region 
between the tagging jets with soft partons. These can recombine into the third-hardest jet which then ends up 
in the rapidity gap. However, the 
$y^*_3$ distribution for \herwig \, and \ds \, is fairly unaffected by the lower $p_{T,j_3}$-cut. The shape 
of the curves stay the same compared to the NLO prediction. The collinear region is well described by the two showers, whereas the rate in the central region is 
too low. \\

A similar effect can be seen when the distance 
parameter $R$ of the anti-$k_T$ algorithm is increased. As an example the right hand side of Figure \ref{fig:07} shows the 
$y_3^*$ distribution for $R=0.7$ instead of the default $R=0.5$. In \pythia, the jet activity in 
the rapidity gap rises compared to the fixed-order NLO prediction since more soft and/or collinear partons are clustered in the third jet. This 
increases the possibility of the third jet to be detected. 
In contrast, \herwig \, and \ds \,
tend to radiate collinearly between the tagging jets and the beam remnant which leads to jets with high $|y_3^*|$. 
This collinear radiation does not affect the shape of $y_3^*$ when lowering the $p_{T,j_3}$-cut, 
it does however change by increasing $R$ (Figure \ref{fig:07}, right). For $R=0.7$, the two \herwig \, showers produce more jets in 
the collinear region than \vbfnlo. 
Since \herwig \, and \ds \, predict the same behavior for the $y_3^*$-variable, one can conclude that the difference between 
\herwig \, and \pythia\, is not caused by wide-angle, soft radiation which is included in \ds. Therefore, truncated shower effects play a minor role. The 
difference between the two \herwig -showers and \pythia \, rather seems to depend on how the available phase space is filled 
with soft and collinear radiation.\\

\begin{figure}[t]
	\centering
	\includegraphics[angle=-90,width=1\textwidth]{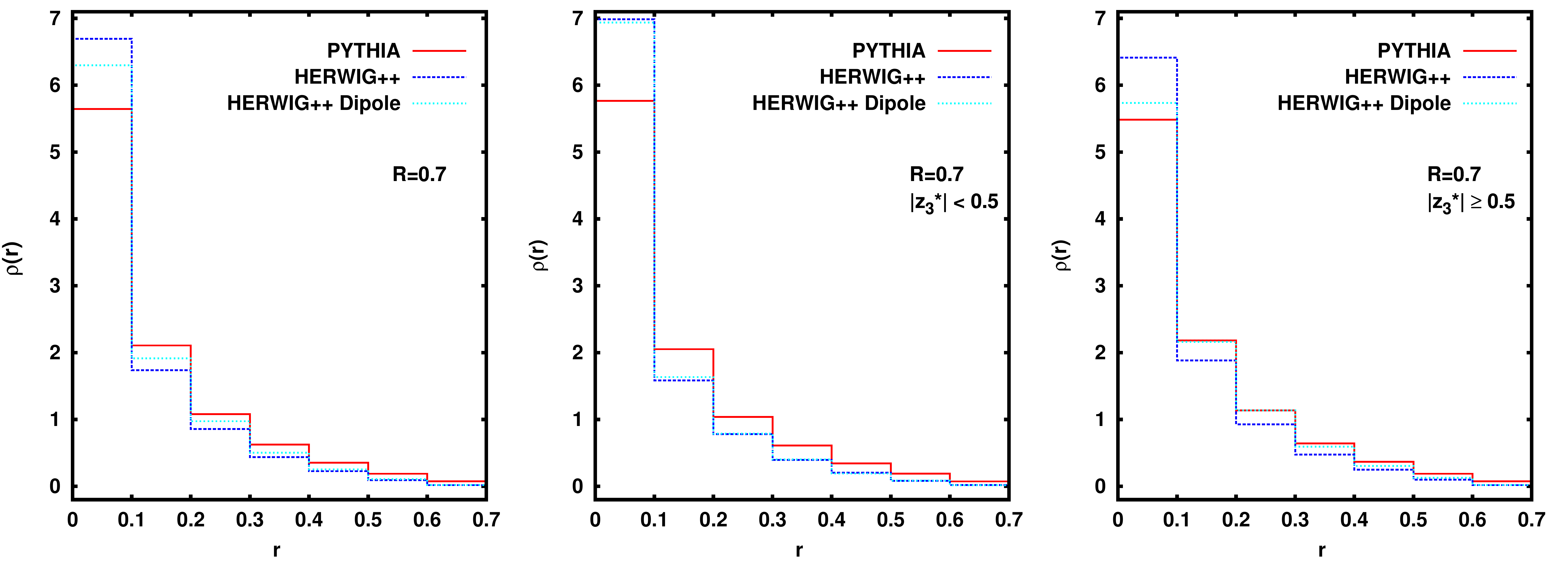}
\caption{ $\rho(r)$ distribution for $R=0.7$ for different regions of the third jet.  
The line styles are assigned as in Figure \protect\ref{fig:same}.}
\label{fig:07ptj3}
\end{figure}

Additionally, the jets obtained with 
\pythia \, are broader than the \herwig \,and \ds \, ones, which is also responsible for the different 
behavior of the three showers in the rapidity gap. This can be seen from the differential jet shape \cite{jetshape} $\rho(r)$ 
of the third jet, which is a measure for the jet energy flow. Following Reference \cite{Aad:2011kq} we define the differential 
jet shape as
\beq
\rho(r)=\frac{1}{\Delta r}\sum_{parton \in j_3}\frac{p_{T,parton}(r-\Delta r/2,r+\Delta r/2)}{p_{T,j_3}}
\eeq
with $r$ ranging between $\frac{\Delta r}{2}$ and $R-\frac{\Delta r}{2}$.
$p_{T,parton}(r_1,r_2)$ denotes the $p_T$ of partons in an annulus 
between radii $r_1$ and $r_2$, i.e. $r_1 \leq r=\sqrt{\left(\phi_{j_3}- 
\phi_{parton}\right)^2 + \left(y_{j_3}- y_{parton}\right)^2} < r_2$. The sum runs over all partons which are recombined into the third jet. 
We use $\Delta r=0.1$ here and  
the normalization assures that $\int_0^R \rho(r) dr =1$. 
In Figure \ref{fig:07ptj3}, the averaged $\rho(r)$ is plotted for the 
third jet with distance parameter $R=0.7$ for different areas of the phase space. 
To distinguish the position of the third jet, we use the variable 
\beq
z_3^*=\frac{y_3^*}{|y_{j_1}-y_{j_2}|}.
\eeq
The tagging jets are localized at $|z^*_3|=0.5$, $|z^*_3|<0.5$ corresponds to the rapidity gap and $|z^*_3|>0.5$ 
to the region between the tagging jets and the beam axis.\\
 
On the left hand side of Figure \ref{fig:07ptj3}, the differential jet shape $\rho(r)$ is plotted in the whole allowed 
phase space. Clearly, \pythia \, produces broader jets than \herwig \, and \ds. The probability to find partons with $r>0.1$ 
which are clustered into the third jet are considerably higher than for the other two parton showers. The middle plot shows $\rho(r)$ 
for a third jet falling in the rapidity gap between the two tagging jets, in the right plot the jet falls into the collinear 
region between the tagging jets and the beam axis. It is noticeable that much of the difference between \herwig \, and \pythia \, 
stems from jets in the central region.
This matches the observations made before and is also correlated on how the available 
phase space for additional radiation is filled: collinear radiation leads to narrow jets, whereas soft, wide-angle radiation 
is rather uncorrelated to the parton where it is radiated off and can therefore broaden the jet. For \pythia, this effect is large for jets 
in the central region (Figure \ref{fig:07ptj3}, middle), whereas the differential jet shape for jets between the tagging 
jets and the beam axis is almost the same for \pythia \, and \ds. In this region, the jet behaves almost like in an inclusive jet 
sample, which is reasonably well described by all three showers \cite{Aad:2011kq}. Compared to \herwig, \ds \, predicts slightly broader jets. This 
can be explained by additional soft radiation due to a low IR cut-off on the Sudakov factor in \ds. By increasing this cut-off, 
agreement in the differential jet shape and the rate of third jet can be obtained.\\

The broadening of the third jet in the central region 
is also the reason why, over a large range of $p_T$, the $p_{T,j_3}$-curve of \pythia \, 
lies below the \herwig \, prediction (see Figure \ref{fig:j3}, left). \pythia \, radiates 
wide-angle partons which are not clustered into the jet and therefore take away part of the original $p_T$ from the parent parton. In \herwig \, many radiated 
partons get clustered along the axis of the parent parton to the third jet as well as additional radiation from the two tagging jets. This can 
be seen in the $p_{T,j_3}$-distribution for large distance parameters $R \geq 0.5$, were it exceeds even the NLO prediction. To see the 
effect of radiation coming from the tagging jets also in \pythia, the distance parameters $R$ has to be increased even more. The 
difference between \herwig \, and \ds \, comes from the different normalization of the two curves. As mentioned before, \ds \, is known 
to radiate more soft partons which can lead to a lower rate of the third jet once a minimum $p_T$ and a maximum rapidity 
threshold on the tagging jets is set. \\

\begin{figure}[t]
	\centering
        \includegraphics[angle=-90,width=1\textwidth]{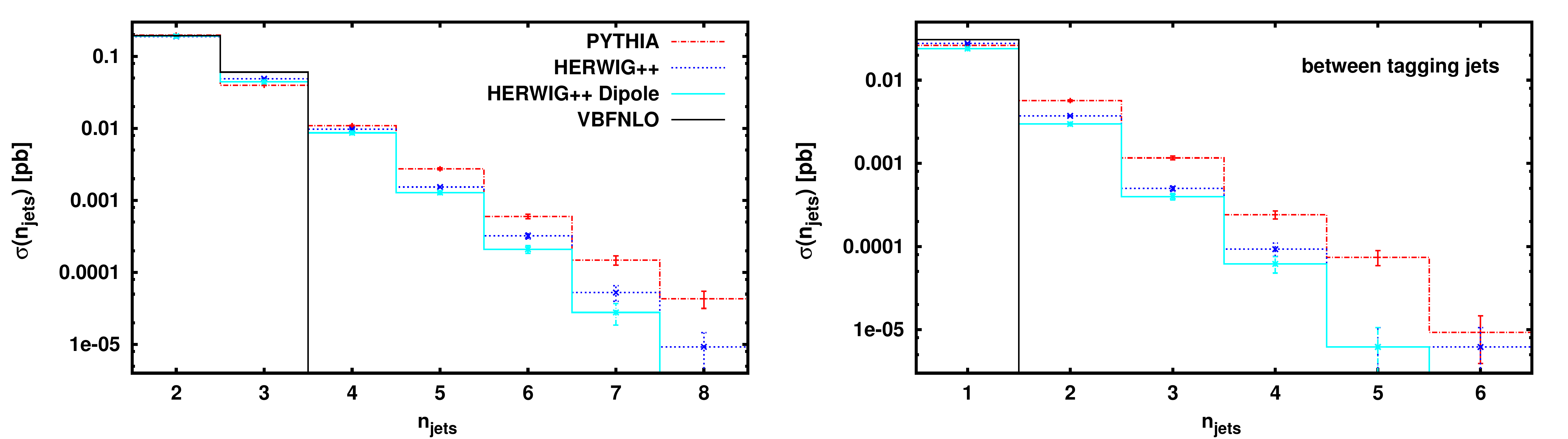}
\caption{Jet multiplicity with the standard cuts (\ref{eq:tag})-(\ref{eq:vbf}) for jets in the whole 
allowed phase space (left) and in the rapidity gap between the tagging jets (right). 
The line styles are assigned as in Figure \protect\ref{fig:same}.}
\label{fig:multipl}
\end{figure}

All this has consequences for CJV techniques, since the multiplicity of jets between the tagging jets 
is quite different for \pythia, \herwig \, and \ds, as shown in Figure \ref{fig:multipl}. The first three jets, which come 
from the hard matrix elements, are reasonably well described by all three showers, whereas additional jets, which 
solely come from the showers, show big differences. \pythia \, radiates off more partons which survive the jet criteria 
(\ref{eq:3rd}, \ref{eq:yj}, \ref{eq:jetlep}), both in the whole allowed phase space and in 
the rapidity gap between the two tagging jets.  
One other important conclusion of this work is that \herwig, as a vetoed angular-ordered shower shows the same behavior as its 
$p_T$-ordered sibling \ds. Therefore, at least for the processes studied here, the effect of truncation can be neglected. 

\section{Conclusions \label{sec:conclusions}}
We implemented $Wjj$ and $Zjj$ production via VBF in the \pbox. The \powheg \, framework allows to interface an NLO calculation with 
parton showers. One basic finding is that, as expected, the shape of the distributions of the two tagging jets and the leptons are 
mostly independent of parton shower effects. Small changes appear in the overall cross sections which arise from 
migration of some events to phase space regions which are not incorporated within the cuts. 
In contrast, the distribution of the third jet, which is only 
LO accurate, is sensitive to the details of the parton shower in use. Dependent on the cuts used, the effect 
on third jet distributions can easily be of the order of $30-40\%$. Since the standard angular-ordered \herwig \, shower and the new 
$p_T$-ordered \herwig-Dipole Shower are in good agreement, we expect the effects of additional wide-angle soft radiation, which 
is missing in the vetoed angular-ordered shower, to be small. However, there exist sizable differences between \pythia \, and \herwig. 
This is due to the fact that \pythia \, predicts broader (third) jets 
than \herwig, especially in the central region of the detector. These stem from soft, wide-angle radiation. 
In \herwig, the third jet tends to be located in the region outside the 
rapidity gap due to additional small-angle radiation. 
The difference between the two \herwig-showers and \pythia \, seems to be caused by the filling of the available phase space for additional 
radiation by the respective shower: \pythia \, tends to fill the rapidity gap between the two tagging jets with 
rather soft partons, while \herwig \, leaves the rapidity gap essentially unaltered and radiates additional partons preferentially in the collinear region 
between the tagging jets and the beam axis.\\

These differences between the three shower predictions reflect remaining uncertainties of available NLO predictions. They are mostly 
present in the distributions of the third jet, since it is only LO accurate, and have to be taken into account when comparing the 
predictions to data.

\section*{Acknowledgments}
\noindent
We thank Stefan Gieseke for many useful discussions concerning the differences between 
\herwig \, and \pythia \, and for carefully reading the manuscript.
F.S. was supported by the "Graduiertenkolleg 1694, 
Elementarteilchenphysik bei h\"ochster Energie und h\"ochster Pr\"azision".
The Feynman 
diagrams have been drawn using the package \textsc{FeynMF} \cite{Ohl:1995kr}.

\end{document}